\documentclass[a4letter, 10 pt, conference]{ieeeconf}  

\IEEEoverridecommandlockouts                              

\usepackage{cite}
\usepackage[hidelinks]{hyperref}
\usepackage{amsmath,amssymb,amsfonts}
\usepackage{algorithmic}
\usepackage{graphicx, subfigure}
\usepackage{textcomp}
\usepackage[colorinlistoftodos,prependcaption,textsize=tiny]{todonotes}
\usepackage{comment}
\usepackage[nolist]{acronym}
\usepackage{gensymb}
\usepackage[utf8]{inputenc}
\usepackage{adjustbox}
\usepackage{array}

\newcolumntype{R}[2]{%
    >{\adjustbox{angle=#1,lap=\width-(#2)}\bgroup}%
    l%
    <{\egroup}%
}
\newcommand*\rot{\multicolumn{1}{R{60}{1.5em}}}

\title{\LARGE \bf
Cyber-Physical Mobility Lab:\\
An Open-Source Platform for Networked and Autonomous Vehicles
}

\author{Maximilian Kloock$^{1}$, Patrick Scheffe$^{1}$, Janis Maczijewski$^{1}$, Alexandru Kampmann$^{1}$, \\Armin Mokhtarian$^{1}$, Stefan Kowalewski$^{1}$ and Bassam Alrifaee$^{1}$
\thanks{$^{1}$ Maximilian Kloock, Patrick Scheffe, Janis Maczijewski, Alexandru Kampmann, Armin Mokhtarian, Stefan Kowalewski and Bassam Alrifaee are with the Chair for Embedded Software, RWTH Aachen University, 52074 Aachen, Germany
        {\tt\{kloock, scheffe, maczijewski, kampmann, mokhtarian, kowalewski, alrifaee\}@embedded.rwth-aachen.de}}
}

\begin{document}

\maketitle
\thispagestyle{empty}
\pagestyle{empty}

\begin{abstract}
This paper introduces our \ac{CPM Lab}. It is an open-source development environment for networked and autonomous vehicles with focus on networked decision-making, trajectory planning, and control. 

The \ac{CPM Lab} hosts 20 physical model-scale vehicles (µCars) which we can seamlessly extend by unlimited simulated vehicles. The code and construction plans are publicly available to enable rebuilding the \ac{CPM Lab}. 

Our four-layered architecture enables the seamless use of the same software in simulations and in experiments without any further adaptions. A \ac{DDS} based middleware allows adapting the number of vehicles during experiments in a seamless manner. The middleware is also responsible for synchronizing all entities following a logical execution time approach to achieve determinism and reproducibility of experiments. This approach makes the \ac{CPM Lab} a unique platform for rapid functional prototyping of networked decision-making algorithms. 

The \ac{CPM Lab} allows researchers as well as students from different disciplines to see their ideas developing into reality. We demonstrate its capabilities using two example experiments. We are working on a remote access to the CPM Lab via a web-interface.

\end{abstract}

\section*{Supplementary material}
A demonstration video of the \ac{CPM Lab} is available at \url{https://youtu.be/PfM0qdzorCc}. 

The code, bill of materials and a construction tutorial is publicly available at \url{https://cpm.embedded.rwth-aachen.de}.  

\acresetall

\section{INTRODUCTION}
\label{sec:introduction}
Testing algorithms for networked and autonomous vehicles is time-consuming and expensive. Full-scale tests of, e.g., decision-making methods require a test track. Tests on public roads may be not eligible. Nowadays, a safety driver has to be in each vehicle to monitor the movement of the vehicle and intervene if required. In addition, one vehicle is not enough to test and evaluate algorithms for networked vehicles. Therefore, multiple vehicles have to be acquired, which increases the cost and logistic overhead. Additionally, the vehicles' software have to be compatible to each other and to the infrastructure, e.g., traffic light communications. As a result, many research institutes have one full-scale test vehicle, but only a few have multiple vehicles for tests of networked algorithms.  

\begin{figure*}[thbp!]
    \centering
    \vspace{1.5mm}
    \includegraphics[width=\textwidth]{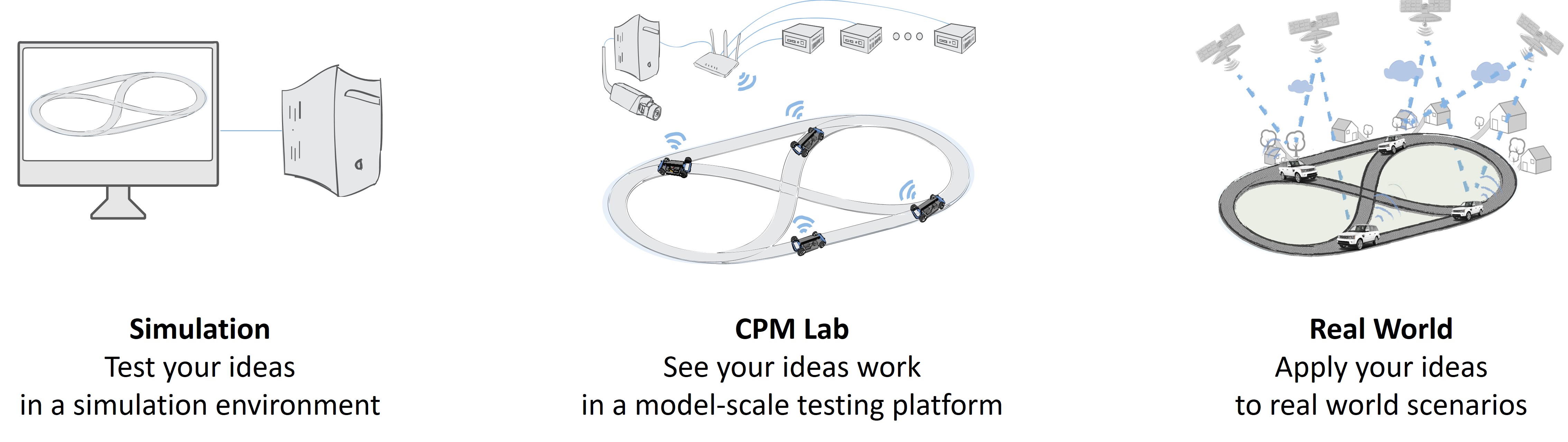}
    \caption{An overview of the development process of simulations (left), \ac{CPM Lab} experiments (middle), and real world experiments (right).}
    \label{fig:sim_lab_real}
\end{figure*}

Because of the shortcomings of full-scale experiments, simulations are the most common way to evaluate algorithms for networked vehicles. Simulations enable concepts like rapid functional prototyping, since changes in the algorithms can be rapidly applied and the results can be seen online. However, simulations abstract from real-world behavior and some aspects may not be included. This results in a big gap between simulations and real-world experiments. In order to mitigate this big gap, we developed the \ac{CPM Lab}. The \ac{CPM Lab} is a model-scale testing platform for networked and autonomous vehicles with focus on decision-making algorithms. The \ac{CPM Lab} simulates inaccuracies due to scale absence, e.g., positioning system inaccuracies, synchronization errors, or communications problems. Hence, the \ac{CPM Lab} reduces the gap between simulations and real-world full-scale experiments. Figure~\ref{fig:sim_lab_real} illustrates the position of the \ac{CPM Lab} in the development and testing process of networked and autonomous vehicles. 

Many testbeds for model-scale autonomous vehicles exist at research institutes. They differ in many aspects, e.g., vehicle hardware, scale, cost, positioning system or communications.
An overview of robots developed in the last decade that cost less than \$300 is given in~\cite{paull2017duckietown}. All robots in this overview are with slip-stick forwards motion, e.g.,~\cite{rubenstein2012kilobot, rubenstein2015aerobot}, or differential wheeled robots, e.g.,~\cite{robinette2009, kernbach2011swarmrobot, cole2012stem, riedo2013thymio, mclurkin2014robot, wilson2016pheeno, paull2017duckietown, stager2017scaled,marquez-sanchez_construction_2019,pickem_robotarium_2017}.
Labs that include vehicles with Ackermann steering geometry are presented in, e.g.,~\cite{liniger2014,gonzales2016,karaman2017,okelly2019,hyldmar2019,bemani_cyber-physical_2020}.
When model-scale vehicles are larger, they typically carry more onboard sensors, e.g., lidar sensors and cameras, and more computation power, but are more costly and need more space to operate, e.g.,~\cite{reiter2017,vedder_low-cost_2018}. Communications between the vehicles include Bluetooth, radio, light signals and WLAN.
\autoref{tab:related_labs} compares closely related labs. The comparison considers the properties: (1) Ackermann steering geometry, (2) capability for autonomous driving, (3) capability for networked driving, (4) open source availability and (5) synchronization of networked control to ensure determinism. Parentheses indicate that the property is either not the focus of the platform or that the property is given with restrictions.
The widely known Duckietown \cite{paull2017duckietown} uses differential drive robots. Communication between the vehicles is only possible through LED signals received by an onboard camera.
The F1TENTH \cite{okelly2019} platform uses the \ac{ROS} platform for communication, so networked testing possibilities are theoretically given. It would need an extension of the architecture, as the focus of the platform lies in autonomous driving. The vehicles are twice as large as our vehicles, so the lab environment for networked research needs to be larger as well.
The characteristic feature of the Prorok Lab \cite{hyldmar2019} is that all computations run on external PCs, and the vehicles only receive actuator signals via radio. The vehicles carry no onboard sensors or considerable computing power.
The Robotarium \cite{pickem_robotarium_2017} runs experiments with differential drive robots. It offers a remote access, so that any registered user can run experiments on swarm robotics.
\begin{table}
    \caption{Recent ground vehicle testing platforms.}
    \centering
    \begin{tabular}{ r | ccccc }
            
            & \rot{Ackermann}
            & \rot{Open source}
            & \rot{Autonomous}
            & \rot{Networked}
            & \rot{Synchronous}
        \\
        \hline
        Duckietown \cite{paull2017duckietown}
            & --
            & \checkmark
            & \checkmark
            & (\checkmark)
            & --
        \\
        Robotarium \cite{pickem_robotarium_2017}
            & --
            & \checkmark
            & \checkmark
            & \checkmark
            & --
        \\
        F1TENTH \cite{okelly2019}
            & \checkmark
            & \checkmark
            & \checkmark
            & \checkmark
            & --
        \\
        Prorok Lab \cite{hyldmar2019}
            & \checkmark
            & \checkmark
            & \checkmark
            & \checkmark
            & --
        \\
        \ac{CPM Lab}
            & \checkmark
            & \checkmark
            & \checkmark
            & \checkmark
            & \checkmark
        \\
    \end{tabular}
    \label{tab:related_labs}
\end{table}

The vision of our \ac{CPM Lab} is to provide an open-source model-scale lab for real-time experiments and rapid-functional prototyping for networked and autonomous vehicles. 
Our focus is on decision-making algorihtms, trajectory planning, and control. The \ac{CPM Lab} simulates components for other functionalities of networked and autonomous vehicles, e.g., perception or prediction. 
In order to provide a testing platform that suits rapid functional prototyping approaches for networked decision-making algorithms, we also provide a simulator of the \ac{CPM Lab} and all its components using the same interfaces as in the \ac{CPM Lab}. 
This enables the seamless use of the same software in simulations and in experiments without any adaptions. 
The \ac{CPM Lab} can test the networked system in a model-, software-, or hardware-in-the-loop scheme, referred to as \ac{XiL}. 

The remainder of this paper is structured as follows. Firstly, Section~\ref{sec:overview} gives a system overview of the \ac{CPM Lab} containing all important modules. Section~\ref{sec:arch} shows the architecture and describes the interaction between all modules of the \ac{CPM Lab}. Section~\ref{sec:examples} presents the operation of the \ac{CPM Lab} in two different scenarios. Finally, Section~\ref{sec:conclusion} concludes the paper.

\section{LAB SETUP}
\label{sec:overview}

\begin{figure}[thbp!]
    \centering
    \includegraphics[width=0.9\columnwidth]{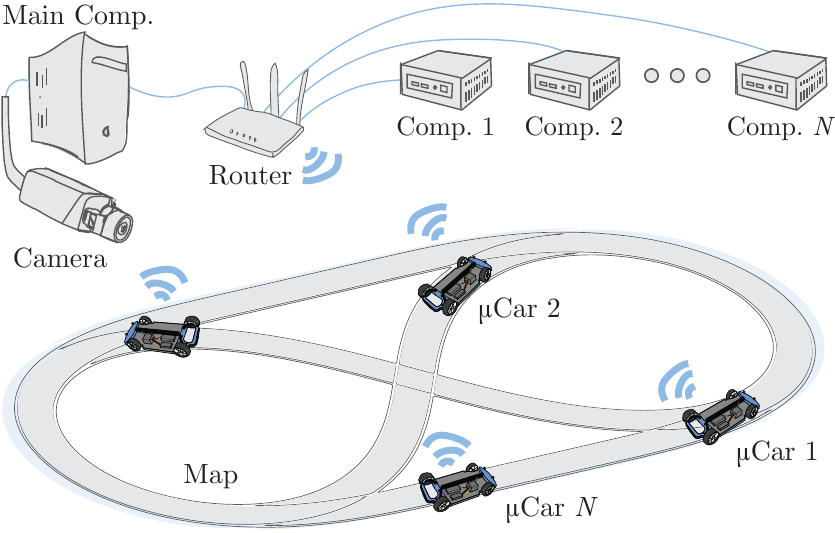}
    \caption{An overview of the \ac{CPM Lab}.}
    \label{fig:overview}
\end{figure}

Figure~\ref{fig:overview} shows a schematic overview of the \ac{CPM Lab}. It consists of 
\begin{enumerate}
  \item 20 model-scale vehicles (µCars), 
  \item a camera for the indoor positioning system,  
  \item external computation devices, 
  \item a main computer to control and monitor experiments, 
  \item a map containing the road structure, and 
  \item a router for the communications. 
\end{enumerate}

\begin{figure}[thbp!]
    \centering
    \includegraphics[width=\columnwidth]{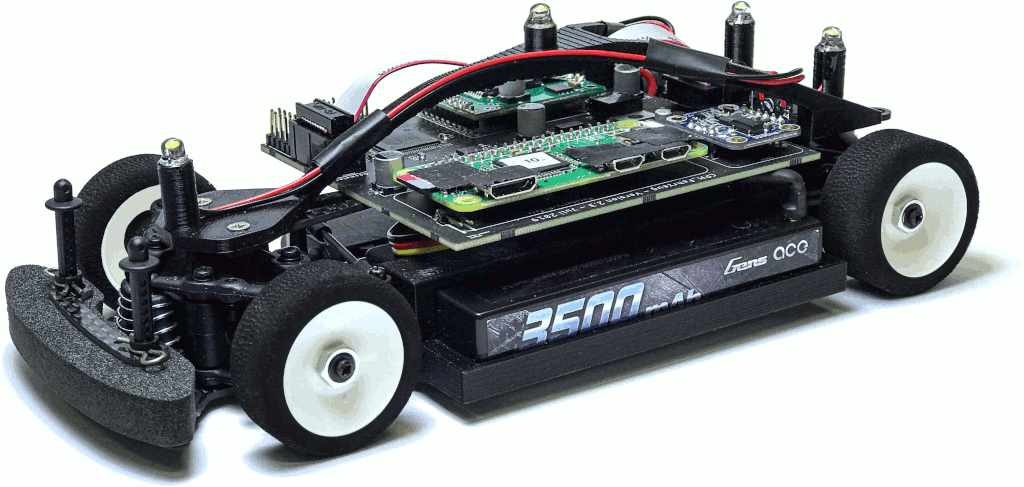}
    \caption{A picture of a µCar.}
    \label{fig:vehicle}
\end{figure}

The 1:18 scale µCars have a length of 220 mm, a width of 107 mm, and a height of 70 mm. The maximum speed is 3.7 m/s. Figure~\ref{fig:vehicle} shows a picture of one µCar. The basis for the µCars is the XRAY M18 Pro LiPo platform~\cite{xray}. 

\begin{figure}[thbp!]
    \centering
    \includegraphics[width=0.75\columnwidth]{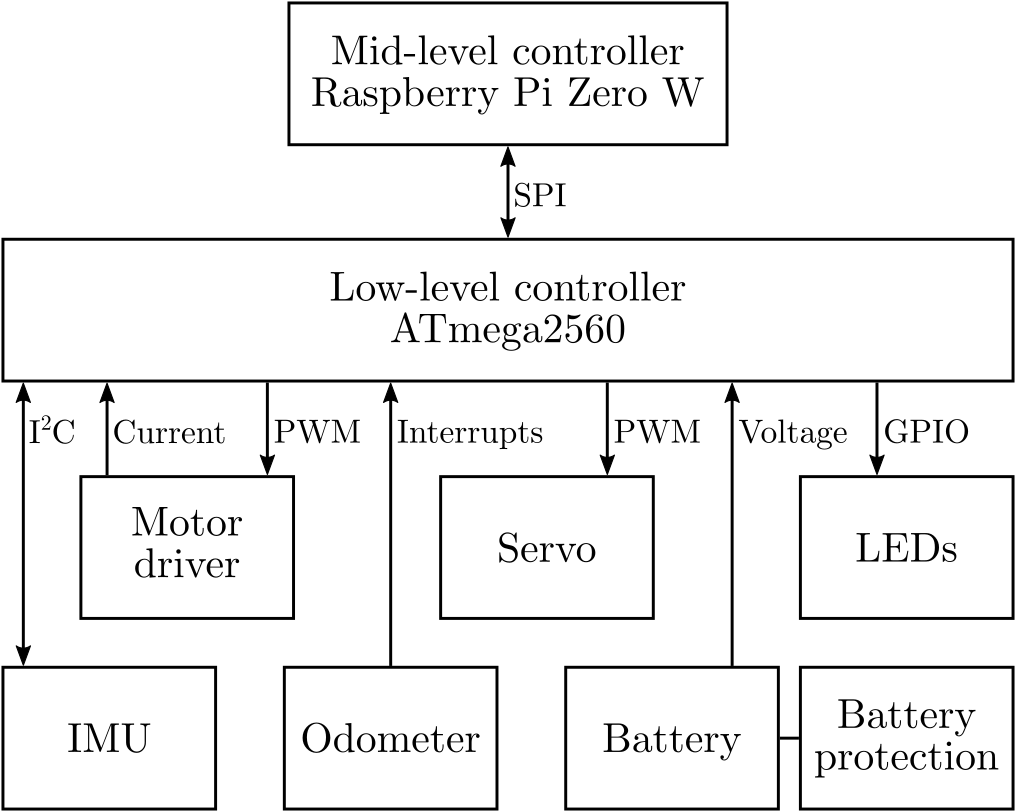}
    \caption{Hardware architecture of the µCar.}
    \label{fig:hardware}
\end{figure}

Figure~\ref{fig:hardware} depicts the hardware architecture of the µCar. 
Each µCar consists of an Atmega 2560 microcontroller, a Raspberry Pi Zero W, an odometer, an IMU and a motor driver. The Atmega 2560 and the Raspberry Pi are used for computations that are described in the architecture in Section~\ref{sec:arch}. The odometer and IMU measure the speed, acceleration, and yaw rate. The odometer is composed of three hall-effect sensors that measure the rotation of a diametrically polarized magnet attached to the motor shaft. The motor driver controls the motor voltage through pulse width modulation. We use servos to steer the µCar. The 3500 mAh batteries allow for five hour runtime. A battery protection inhibits the battery of discharging below a threshold to prevent the battery from damage. The scale of the µCars approximate the scale of Volkswagen Golf 7 sized vehicles. We described in~\cite{scheffe2020} the µCar hardware in detail. 

We developed a vision-based \ac{IPS} that computes the positions and orientations (poses) of the µCars. In order to keep the costs and computation requirements of the µCars low, the poses are computed externally on the main computer. The \ac{IPS} consists of a Basler acA2040 grayscale module camera that is mounted 3.3m above the track and LEDs attached to each µCar.  Each µCar is equipped with four LEDs, see Figure~\ref{fig:vehicle}. The three outer LEDs are used to determine the pose of the µCar. In order to map poses to µCars, the \ac{IPS} also identifies the µCars using the fourth LED. The identification LEDs flash in unique frequencies. The \ac{IPS} identifies LEDs using a 50Hz stream from the camera. We set a low exposure time to have a high contrast of the LEDs to the ambient light. By this, the LED spots are clearly identified through image processing. 
The three outer LEDs build a non-equilateral triangle. By this, the poses can be computed unambiguously. The poses are computed for the center of mass of the µCars. Please note that the \ac{IPS} provides all coordinates in a global form, i.e., the vehicles receive their coordinates in a global coordinate system. Our \ac{IPS} is described in~\cite{kloock2020} in more detail. 

In order to enable rapid functional prototyping, we provide external computation entities to each µCar. These external computation entities are Intel NUCs, equipped with i5 processors and 16 GB of RAM each. For simulation of networked computations, each µCar is logically connected to an Intel NUC. Each Intel NUC only performs computations for one µCar. 

We constructed the map to fit the µCars' dynamics and our space requirements. Due to the continuity of the change of the steering angle, the roadway should be two times continuous differentiable~\cite{meek1992clothoid}. With respect to the maximum steering angle of the µCars the maximum curvature of the road is limited. For space reasons, the map is limited to 4 m x 4.5 m. To keep the space requirements low, the lanes are narrow but fit to the width of the µCars. The roads are for visualization only and are not detected by any mechanism of the \ac{CPM Lab}. The digital representation of the map, nevertheless, is used, e.g., for decision-making. 

\begin{figure*}[thbp!]
    \centering
    \includegraphics[width=0.86\textwidth]{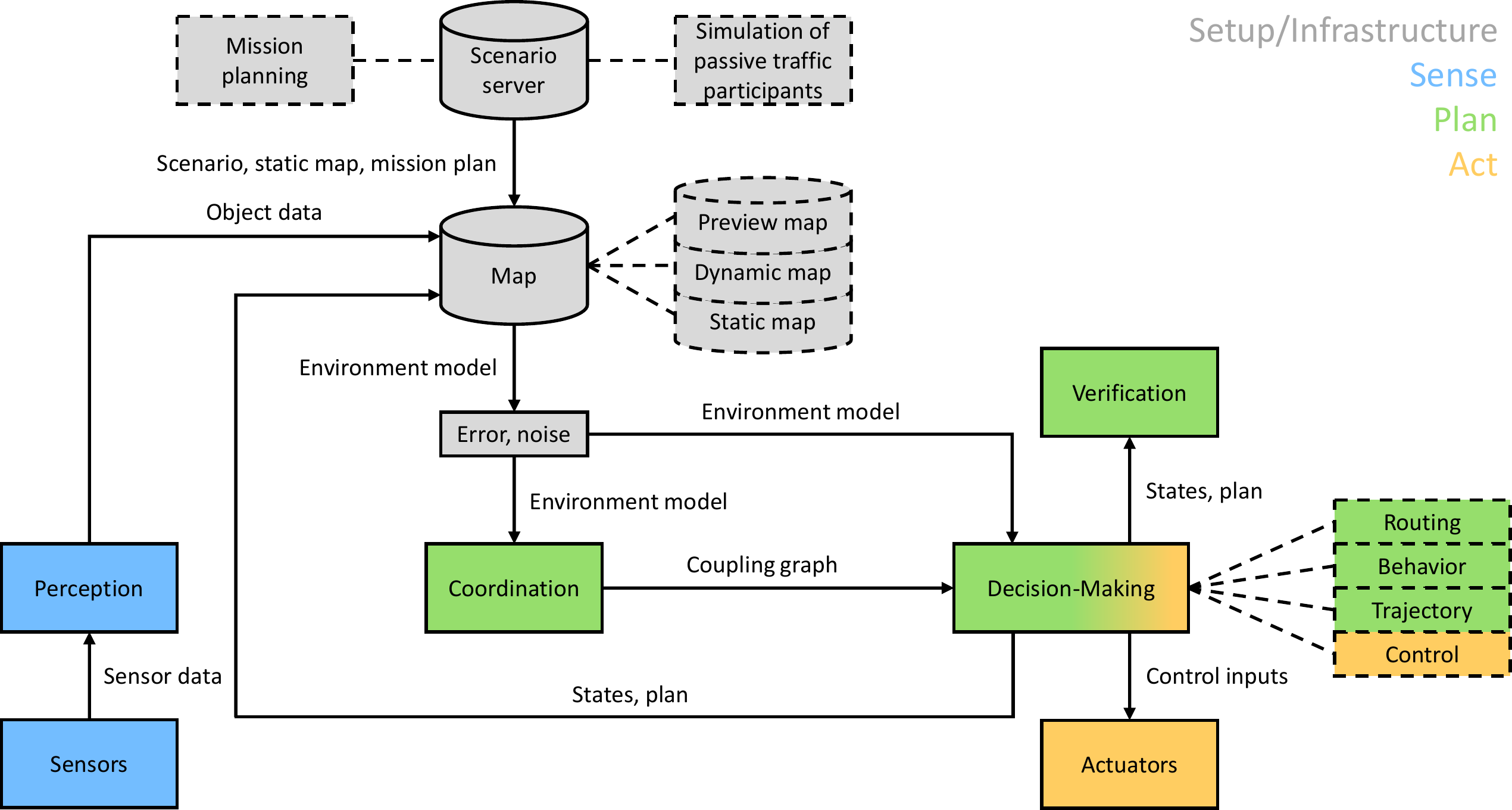}
    \caption{Framework concept. Colors illustrate the logical affiliation. Grey, blue, green and yellow denote infrastructure, \textit{Sense}, \textit{Plan} and \textit{Act}, respectively.}
    \label{fig:framework}
\end{figure*}

Figure~\ref{fig:framework} shows the framework architecture of the \ac{CPM Lab}. It follows the \textit{Sense}, \textit{Plan}, \textit{Act} scheme, including infrastructure functionalities. 

\paragraph{Infrastructure}
The infrastructure part provides a database of scenarios, called scenario sever. Scenarios include mission plans and the simulation of non-automated traffic participants. Scenarios are defined in the CommonRoad scenario format of~\cite{althoff2017commonroad}. The scenario data are stored in the map. The map is used as a database at runtime and includes static data like the road network, dynamic data like the positions of traffic participants, and preview data of planning. The road network is stored in the Lanelets road format of~\cite{bender2014lanelets}. In order to simulate real environments, the environment model can be affected by artificial errors and noises in different intensities, e.g., to simulate positioning errors or communications delays. 

\paragraph{Sense}
Each µCar consists of an \ac{IMU} and an odometer. The camera externally computes the poses, i.e., positions and orientations of all µCars and communicate them to all µCars. This simulates GNSS. Please note that the focus of the \ac{CPM Lab} is on decision-making, trajectory planning, and control. Hence, the \ac{CPM Lab} simulates most of the sensors that are used in networked and autonomous vehicles. 

\paragraph{Plan}
The planning part consists of the modules coordination, decision-making, and verification. The coordination module determines the coupling of the µCars for the decision-making. The decision-making consists of the submodules routing, behavior, trajectory, and control. The routing submodule plans the route from a start position to an end position. The behavior submodule plans the behavior of the µCar and the trajectory submodule computes trajectories. Before the trajectories are applied on the µCar, they are verified to ensure safety aspects, e.g., collision-freeness. Our work in~\cite{volker2019verification} is an example of verification, while the works in~\cite{alrifaee2016coordinated,kloock2019distributed2,kloock2019networked} are examples of decision-making. The \ac{CPM Lab} is able to execute the decision-making of multiple µCars centralized, or distributed in a parallel, sequential, or hybrid manner. Section~\ref{sec:arch} gives more details about the decision-making architecture and Section~\ref{sec:examples} presents two example experiments. 

\paragraph{Act}
The acting part consists of the decision-making submodule control and the physical actuators. The submodule control uses the planned trajectory as an input and computes corresponding control inputs, i.e., motor voltage and steering angle. The resulting commands are executed by the motor driver and servo. 
 
\section{ARCHITECTURE} 
\label{sec:arch}
\begin{figure}[thbp!]
    \centering
    \includegraphics[width=\columnwidth]{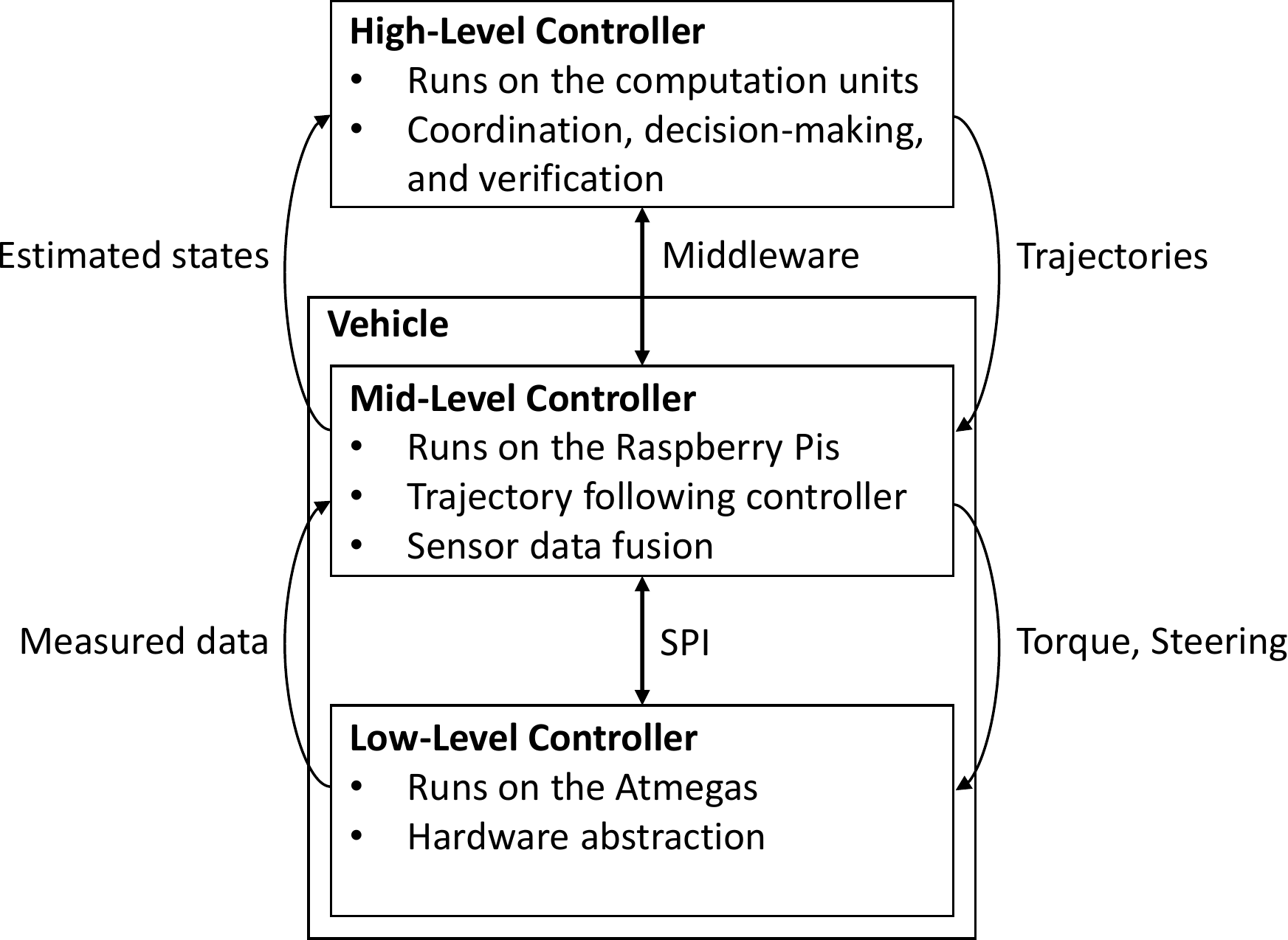}
    \caption{The architecture of the \ac{CPM Lab} is divided into \ac{HLC}, middleware, \ac{MLC}, and \ac{LLC}. The architecture is placed on external computation devices, Raspberry Pis, and ATmega microcontrollers.}
    \label{fig:arch}
\end{figure}

Figure~\ref{fig:arch} illustrates the architecture of the \ac{CPM Lab}. It consists of \acp{HLC} placed on the external computation devices, a middleware, \acp{MLC} placed on the Raspberry Pis, and \acp{LLC} placed on the ATmega microcontrollers. The Raspberry Pis and ATmega microcontrollers are placed on the µCars. The external computation devices are logically connected to the µCars. In the following, we explain all modules of the architecture in detail. 

\subsection{\acf{HLC}}
The \acp{HLC} run on the external computation devices. They are not placed on the µCars due to  space and weight requirements. The \acp{HLC} are responsible for the modules coordination, decision-making and verification, see Figure~\ref{fig:framework}. The \acp{HLC} send trajectories to the \acp{MLC} and receive the fused poses of the µCars from the \acp{MLC}. Depending on the mode of operation, the \ac{HLC} couplings are different. In the following we describe the centralized and distributed computation modes of the \ac{CPM Lab}. 

\subsubsection{Centralized Computation}
For centralized computation, only one \ac{HLC} is needed. This \ac{HLC} is logically connected to all µCars and plans trajectories depending on all µCar states, objectives, and constraints. The output of the central \ac{HLC} consists of one trajectory for each vehicle. Figure~\ref{fig:arch_dist}(a) depicts the architecture for centralized planning. 

\subsubsection{Distributed Computation} 
\begin{figure}[thbp!]
    \centering
    \subfigure[Architecture of centralized planning.]{\includegraphics[width=0.49\columnwidth]{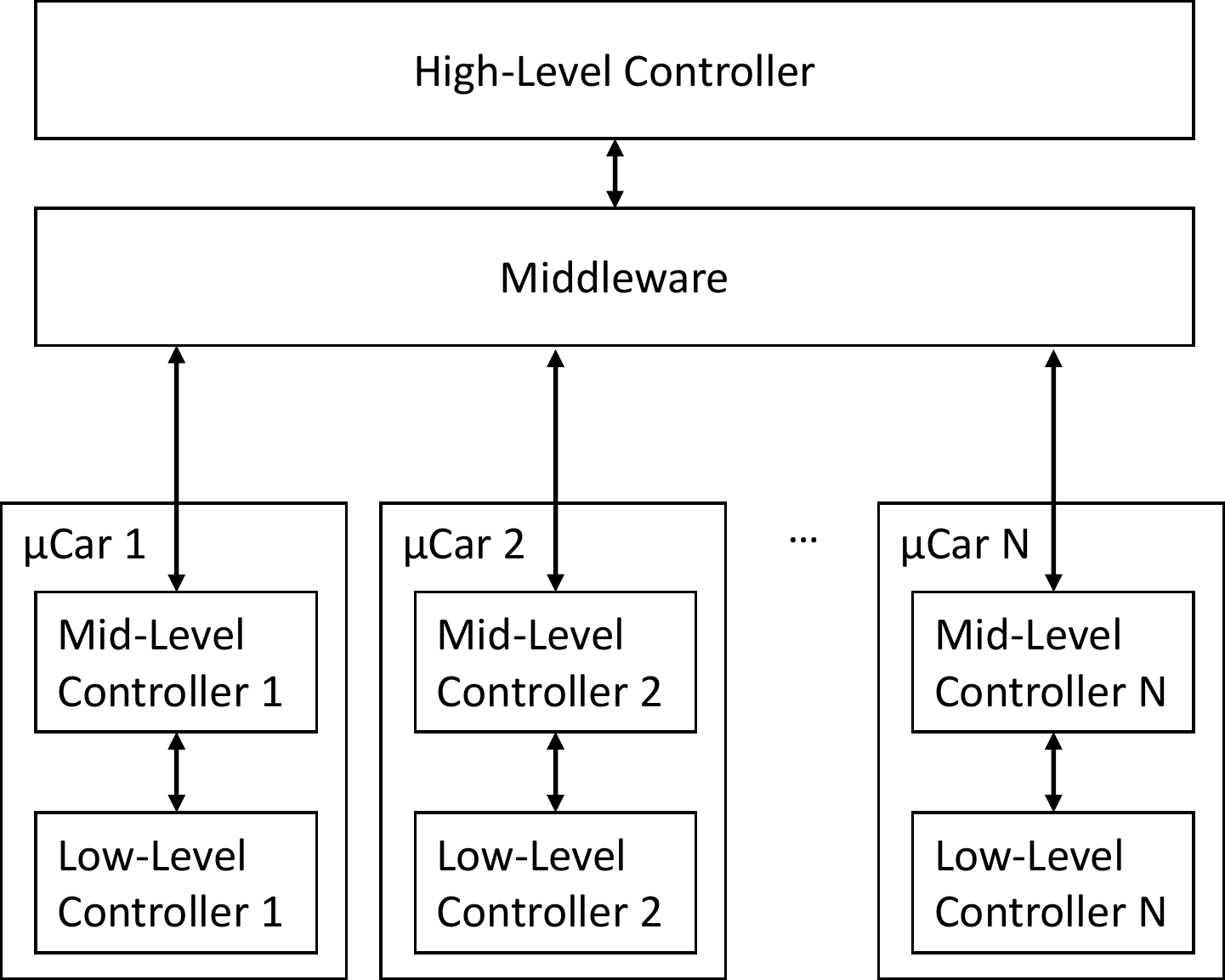}}	
    \subfigure[Architecture of distributed planning.]{\includegraphics[width=0.49\columnwidth]{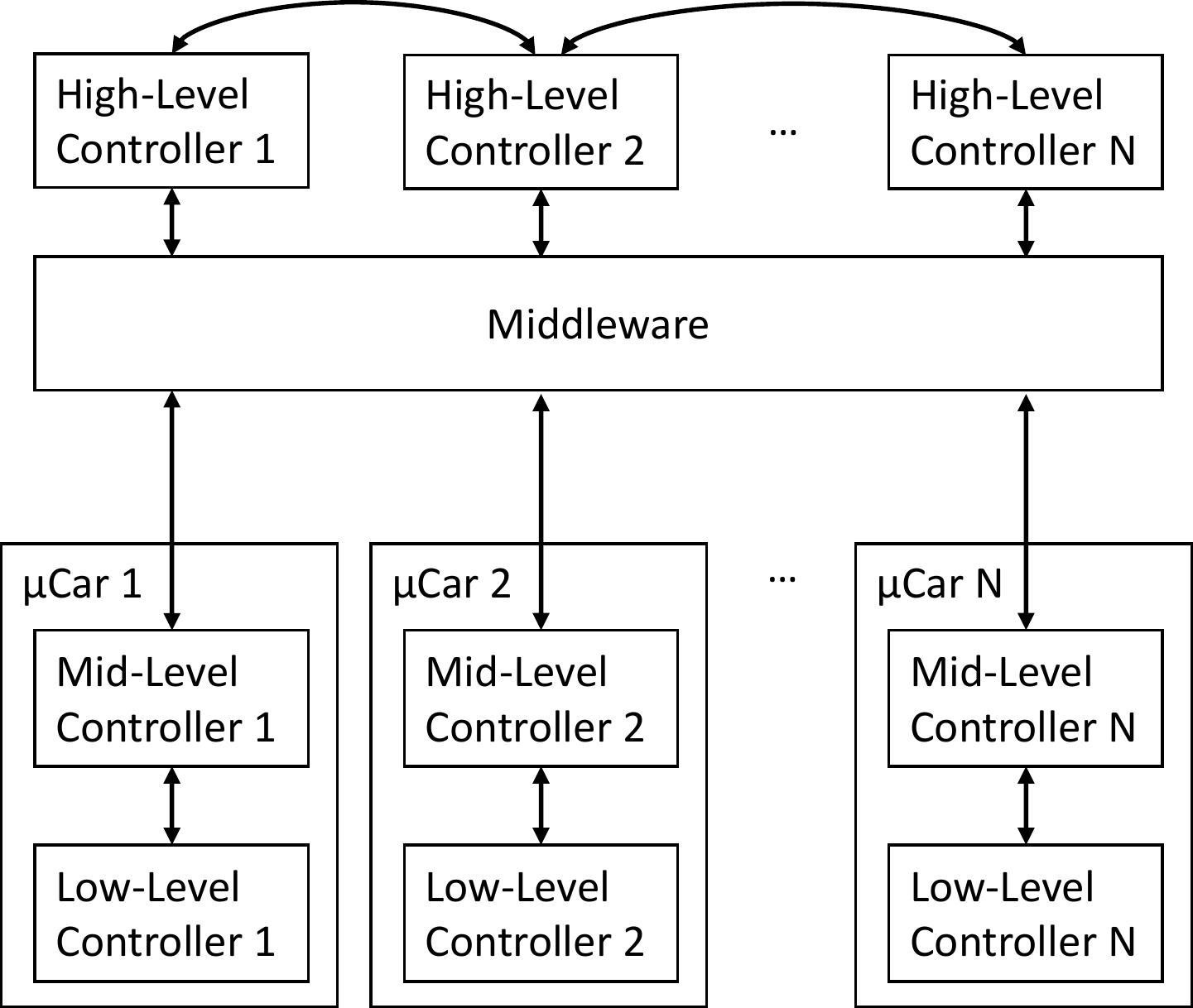}}	
    \caption{The architectures for centralized and distributed planning. In centralized planning, there is only one \ac{HLC} for all µCars.} 
    \label{fig:arch_dist}
\end{figure}

In distributed computation mode, each vehicle is logically connected to an unique \ac{HLC}, i.e., there is a one-to-one mapping for \ac{HLC} to vehicle. This mode represents a system in which each µCar on the road plans its own trajectory without a central coordinator. The trajectory planning requires some cooperation between the \acp{HLC}. Figure~\ref{fig:arch_dist}(b) depicts the architecture for distributed planning. 

\subsection{Middleware}
The middleware runs on the external computation devices and on the Raspberry Pis on the µCars. It synchronizes the computations of the \acp{HLC} and the \acp{MLC} using the concepts of the logical execution time approach~\cite{henzinger2001giotto}. 

The middleware is based on the \ac{DDS}, which is a standardized protocol for decentralized communications in distributed systems based on the publish-subscribe pattern~\cite{pardo2003omg}.
Besides being used in safety-critical systems, such as medical devices and air traffic control~\cite{omg2019whois}, DDS is also entering the automotive domain as part of the upcoming AUTOSAR Adaptive platform~\cite{furst2016autosar}.
The protocol offers a variety of configurable Quality-of-Service (QoS) parameters, including dependable or best-effort communications. 
In contrast to the widespread \ac{ROS}~\cite{quigley2009ros}, DDS does not rely on a designated entity for service discovery or binding, which makes the resulting architecture more robust. 
However, in contrast to \ac{DDS} \ac{ROS} comes along with many packages and tools to support the development process. 

At its core, DDS  uses the User Datagram Protocol (UDP), which leads to lower communications latencies than middlewares based on the Transmission Control Protocol (TCP), such as \ac{ROS} or Message Queuing Telemetry Transport (MQTT)~\cite{mqtt}. 
This is because UDP does not require an acknowledgment for each packet. We do not rely on retransmitting the dropped packets, because the data is time-critical and become obsolete when a packet is dropped. 
Through the use of \ac{DDS}, a variable number of µCars can be part of experiments, without having to adapt the underlying communications architecture.
Additionally, the \ac{CPM Lab} architecture becomes more adaptable for extensions through the dynamic coupling of components in the communications architecture.
Various commercial and open source implementations of DDS are available \footnote{http://www.eprosima.com}$^{,}$\footnote{http://www.opendds.org}. We use the RTI Connext DDS implementation\footnote{https://www.rti.com/products/dds-standard}.

\subsection{\acf{MLC}}
The \acp{MLC} run on Raspberry Pis which are mounted on the µCars. 
The \acp{MLC} provide two modes of operation: direct control and trajectory following.
In direct control, the \acp{MLC} receive commands of torque and steering angle from the \acp{HLC}. 
In trajectory following, the \acp{MLC} receive trajectory nodes of the form $[t_i,x_i,y_i,v_{x,i},v_{y,i}]$, where $t_i \in \mathbb{R}^{+}$ represents the time at which µCar $i \in \mathbb{N}$ should be at position $[x_i, y_i] \in \mathbb{R}\text{x}\mathbb{R}$ with velocity $[v_{x,i},v_{y,i}] \in \mathbb{R}\text{x}\mathbb{R}$ in $x$ and $y$ direction, respectively. 
The continuous reference trajectory $[x_{ref}(t), y_{ref}(t)]$ is constructed using Cubic Hermite spline interpolation, which interpolates between the trajectory nodes.
The use of Hermite interpolation allows the addition of trajectory nodes in real time without affecting the interpolation between previous nodes.
The \acp{MLC} implement trajectory following controllers based on \ac{MPC}.
The \acp{MLC} perform sensor fusions and use the fused poses of the on-board odometers and \acp{IMU} and from the \ac{IPS} via wireless communications. The computed torques and steering angles are communicated to the \acp{LLC}. 

\subsection{\acf{LLC}}
The \acp{LLC} run on ATmega 2560 microcontrollers on the µCars. 
They act as a hardware abstraction layer and sample the on-board sensors, convert the sensor signals into data compatible to the \acp{MLC}, and send the sensor data to the \acp{MLC}. 
The \acp{LLC} apply the torque and steering angle given by the \acp{MLC} to the actuators and convert the control inputs into signals compatible with the µCars' hardware.

\section{EXPERIMENTS}
\label{sec:examples}
We present the use cases of the \ac{CPM Lab} in multiple example experiments. The experiments are implemented in different programming languages. First, we show one experiment with different amount of µCars and different mixtures of real and simulated µCars. Afterwards, we evaluate the distributed planning in the \ac{CPM Lab}. Videos of the presented experiments are shown in the demonstration video available on https://youtu.be/JWOyf19-nKg.

\subsection{Real and Simulated µCars} 
\begin{figure}[tbp!]
    \centering
    \vspace{1.5mm}
    \subfigure[Physical test track in the \ac{CPM Lab}.]{\includegraphics[width=0.8\columnwidth]{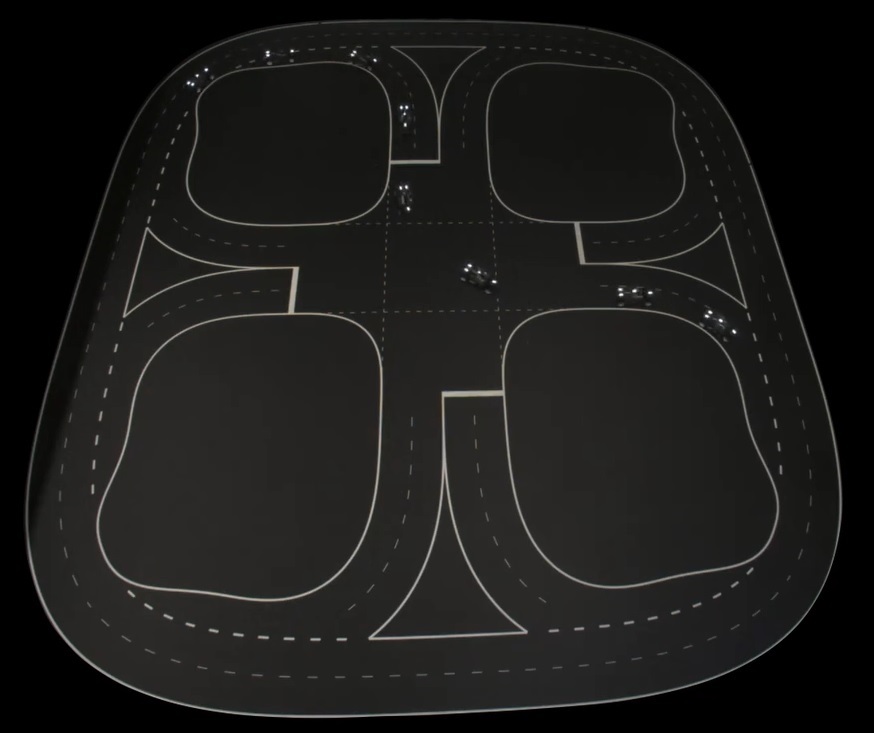}}
    \subfigure[Visualization of the test track.]{\includegraphics[width=0.8\columnwidth]{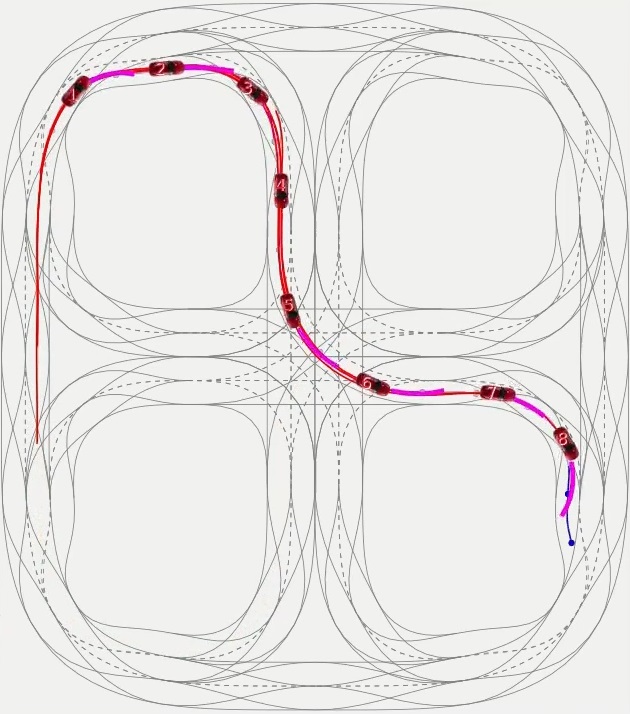}}
    \caption{The test track of the \ac{CPM Lab} with a platoon consisting of 8 µCars. The blue lines represent the µCars' prediction horizon in the \acp{HLC}, the purple lines represent the control horizon in the \acp{MLC}, and the red lines represent the driven path of the µCars.}
    \label{fig:platoon8}
\end{figure}

For the evaluation of real and simulated µCars, we start with a platoon scenario. The platoon scenario consists of a leader µCare and an arbitrary number of following µCars. The leading µCar plans its trajectory given the path and a reference speed. The following µCars adapt the trajectory used by the leading µCar in order to follow it. This experiment is implemented in MATLAB.  Figure~\ref{fig:platoon8} shows eight µCars on the test track in the \ac{CPM Lab}. Figure~\ref{fig:platoon8}(a) shows the view on the physical test track and Figure~\ref{fig:platoon8}(b) shows the visualization. 

\begin{figure}[tbp!]
    \centering
    \vspace{1.5mm}
    \subfigure[Physical test track in the \ac{CPM Lab}.]{\includegraphics[width=0.8\columnwidth]{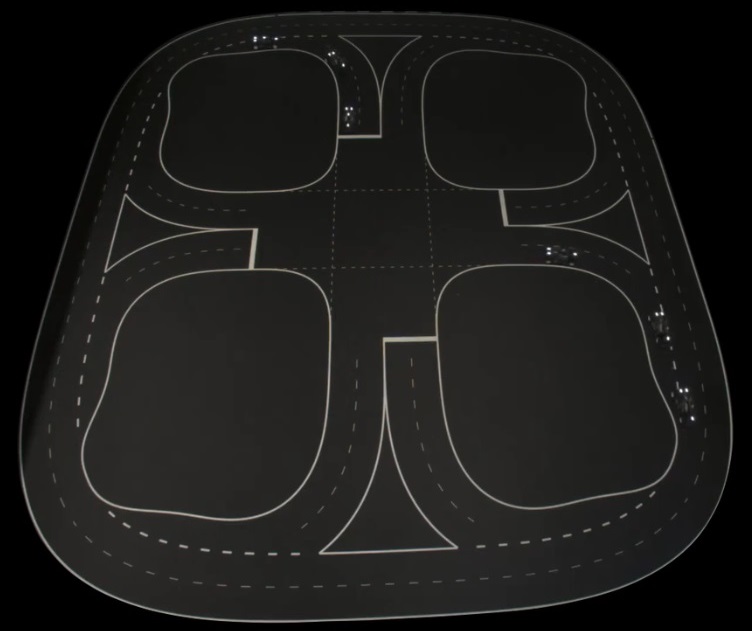}}
    \subfigure[Visualization of the test track.]{\includegraphics[width=0.8\columnwidth]{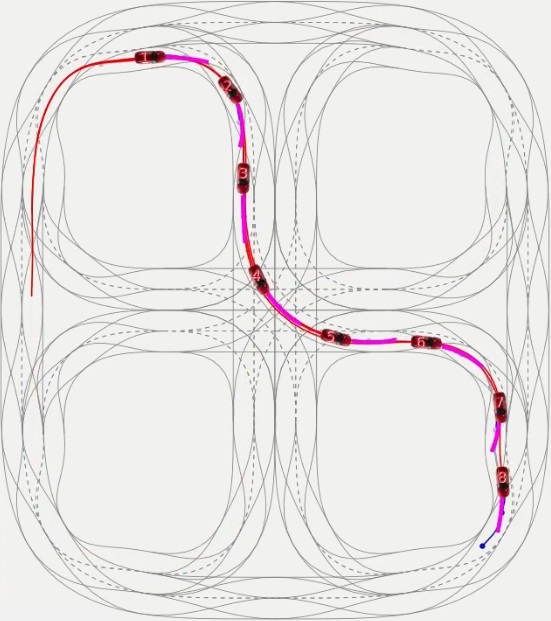}}
    \caption{The test track of the \ac{CPM Lab} with a platoon consisting of 6 real µCars and 2 simulated µCars. The blue lines represent the µCars' prediction horizon in the \acp{HLC}, the purple lines represent the control horizon in the \acp{MLC}, and the red lines represent the driven path of the µCars.}
    \label{fig:platoon6}
\end{figure}

We replaced two real µCars with two simulated µCars and evaluated the platoon scenario again. Figure~\ref{fig:platoon6} shows the test track in the \ac{CPM Lab}.  Figure~\ref{fig:platoon6}(a) shows the physical track with 6 real µCars and  Figure~\ref{fig:platoon6}(b) shows the visualization. The visualization also shows the simulated µCars, since the \ac{CPM Lab} does not differ between those. Therefore, the real µCars behave as if all µCars were driving on the physical track. No modifications of the source code are required, even for using only simulated µCars. Therefore, the implementation for decision-making can be developed in simulations without the need for any additional hardware than the main computer, see Section~\ref{sec:overview}. The task of the main computer can be done by any desktop or laptop computer. 

\subsection{Distributed Computations} 
\begin{figure}[tbp!]
    \centering
    \vspace{1.5mm}
    \subfigure[Physical test track in the \ac{CPM Lab}.]{\includegraphics[width=0.8\columnwidth]{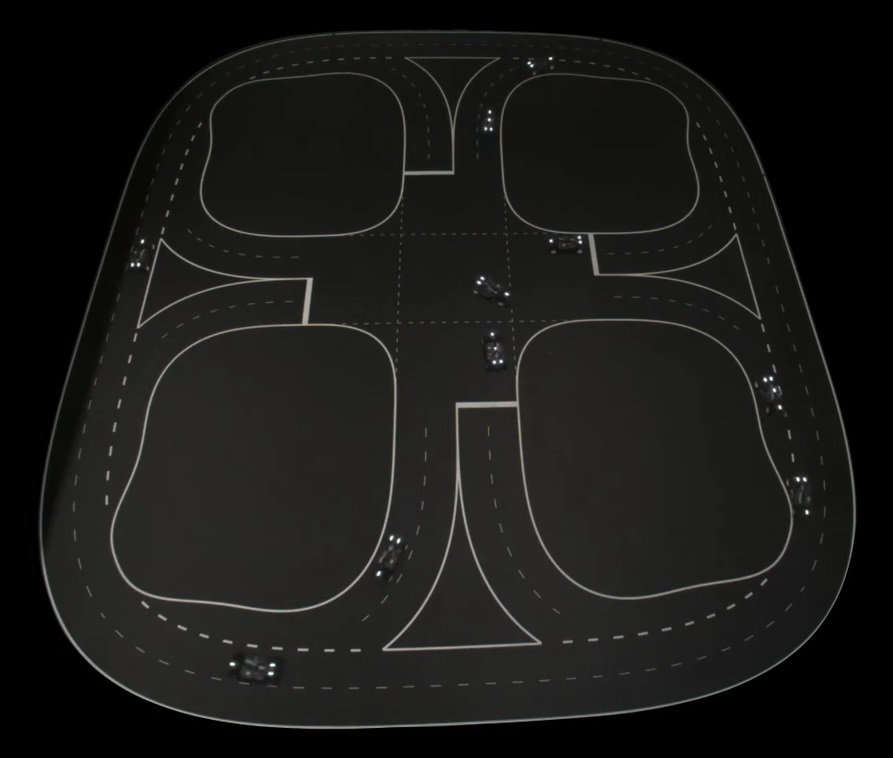}}
    \subfigure[Visualization of the test track.]{\includegraphics[width=0.8\columnwidth]{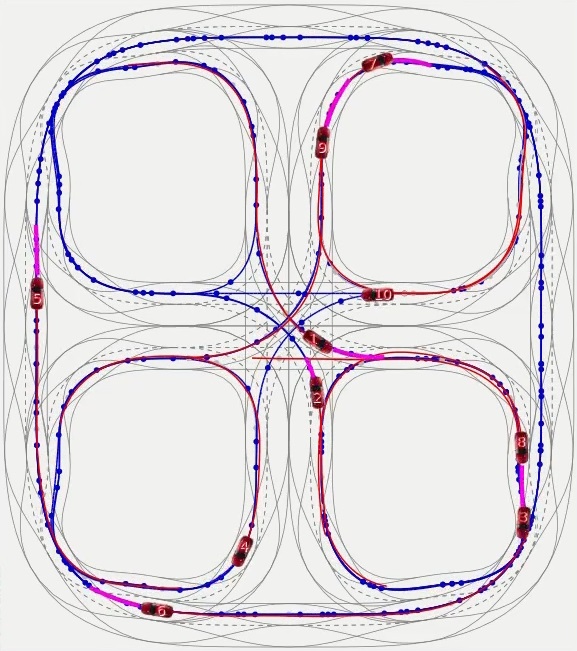}}
    \caption{The test track of the \ac{CPM Lab} with 8 µCars driving in our intersection example. The blue lines represent the µCars' prediction horizon in the \acp{HLC}, the purple lines represent the control horizon in the \acp{MLC}, and the red lines represent the driven path of the µCars.}
    \label{fig:intersection}
\end{figure}

For the evaluation of distributed computations in the \ac{CPM Lab}, we used an intersection scenario. In the intersection scenario, µCars plan their trajectories on the path given a reference speed. The path is chosen randomly at each junction in order to balance left and right moves of the µCars and to simulate real traffic. Then, µCars avoid collisions using a priority-based scheme as proposed in~\cite{alrifaee2017networked}. We statically prioritized the µCars by defining unique µCar IDs. The lower the ID of a µCar the more important it is in the trajectory planning. µCars ignore collisions with µCars of higher ID. Therefore, the responsibility of collision avoidance is at µCars with higher IDs. µCars with higher IDs reduce their speed to avoid collisions with µCars with lower IDs. After avoiding a collision, the µCars accelerate to their reference speed. Please note that in this scenario µCars avoid collisions by speed reduction only. The path is not adapted for collision avoidance. This experiment is implemented in C++. Figure~\ref{fig:intersection} shows the test track of the \ac{CPM Lab} with 8 µCars in the intersection scenario. 

\subsection{Infrastructure Stress-Test}
\begin{figure}[tbp!]
    \centering
    \vspace{1.5mm}
    \subfigure[Physical test track in the \ac{CPM Lab}.]{\includegraphics[width=0.8\columnwidth]{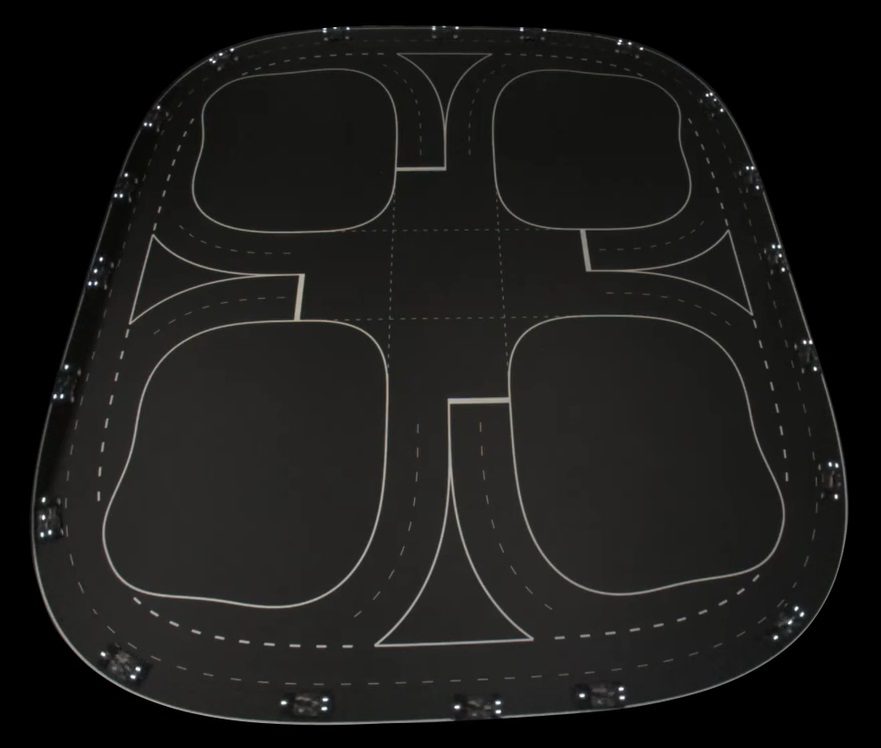}}
    \subfigure[Visualization of the test track.]{\includegraphics[width=0.8\columnwidth]{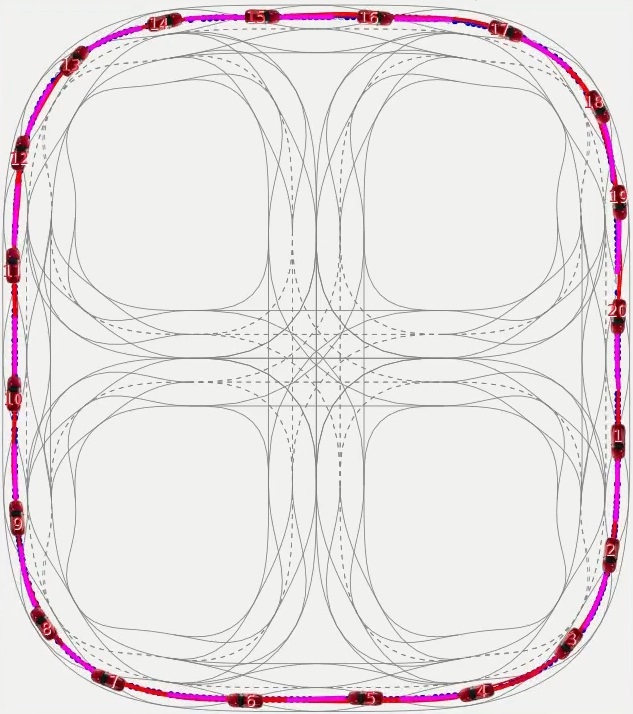}}
    \caption{The test track of the \ac{CPM Lab} with 18 µCars driving in our circle example. The blue lines represent the µCars' prediction horizon in the \acp{HLC}, the purple lines represent the control horizon in the \acp{MLC}, and the red lines represent the driven path of the µCars.}
    \label{fig:circle}
\end{figure}

For testing our infrastructure, including the WiFi communications between the \acp{HLC} and µCars at high load, we placed 18 µCars on the map and let them drive on the outer circle with a fixed speed. Therefore, no collision avoidance is required. The \ac{HLC} sent trajectories for the µCars in a frequency of 100 ms. Figure~\ref{fig:circle}(a) shows the real test track of the \ac{CPM Lab} and Figure~\ref{fig:circle}(b) shows the visualization. The infrastructure of the \ac{CPM Lab} is able to handle 18 µCars on the track with still enough resources to perform tasks in real time. Please note that this infrastructure test only evaluates the \ac{CPM Lab} infrastructure and not a specific decision-making algorithm, since the decision-making is static in this test. The purpose is to evaluate the \ac{CPM Lab} at its limits of 18 µCars. Currently, we are assembling 2 more µCars to complete the fleet of 20 µCars. 

\section{CONCLUSION} 
\label{sec:conclusion}
This paper presented the \ac{CPM Lab}, a seamless development environment for networked and autonomous vehicles with focus on decision-making, trajectory planning, and control. 
We presented two example experiments in our \ac{CPM Lab}. The experiments differ in their programming language and computation architecture. 

Furthermore, we presented our four-layered architecture that enables the use of the same software in simulations and experiments without any adaptions. 
Our middleware allows to adapt the number of vehicles during experiments and simulations. 
The \ac{CPM Lab} can extend experiments with the 20 model-scale vehicles by unlimited additional simulated vehicles. 
We developed an architecture of \ac{HLC}, \ac{MLC}, \ac{LLC}, and middleware to apply new trajectories in the vehicles deterministically and synchronously in a logical execution time approach. Furthermore, we developed the vehicles based on a model-scale RC platform and an \ac{IPS} that computes the poses of the vehicles on the map. 
We used the \ac{CPM Lab} in two practical courses in different study programs with 30 students each. 

\subsection{Outlook}
We are developing a remote-access to the \ac{CPM Lab}. This will allow researchers and students to use the \ac{CPM Lab} without personal presence. 
For more convenient experiments, we will implement our method from~\cite{kloock2019distributed} in the \ac{CPM Lab} to automatically drive the vehicles to their starting poses of experiments. Furthermore, we will apply methods of our previous and future research in the \ac{CPM Lab}.

\section*{ACKNOWLEDGMENT}
This research is supported by the Deutsche Forschungsgemeinschaft (DFG, German Research Foundation) within the Priority Program SPP 1835 “Cooperative Interacting Automobiles” and the Post Graduate Program GRK 1856 “Integrated Energy Supply Modules for Roadbound E-Mobility”.

\bibliographystyle{IEEEtran}
\bibliography{root}

\begin{thebibliography}{10}
\providecommand{\url}[1]{#1}
\csname url@samestyle\endcsname
\providecommand{\newblock}{\relax}
\providecommand{\bibinfo}[2]{#2}
\providecommand{\BIBentrySTDinterwordspacing}{\spaceskip=0pt\relax}
\providecommand{\BIBentryALTinterwordstretchfactor}{4}
\providecommand{\BIBentryALTinterwordspacing}{\spaceskip=\fontdimen2\font plus
\BIBentryALTinterwordstretchfactor\fontdimen3\font minus
  \fontdimen4\font\relax}
\providecommand{\BIBforeignlanguage}[2]{{%
\expandafter\ifx\csname l@#1\endcsname\relax
\typeout{** WARNING: IEEEtran.bst: No hyphenation pattern has been}%
\typeout{** loaded for the language `#1'. Using the pattern for}%
\typeout{** the default language instead.}%
\else
\language=\csname l@#1\endcsname
\fi
#2}}
\providecommand{\BIBdecl}{\relax}
\BIBdecl

\bibitem{paull2017duckietown}
L.~Paull, J.~Tani, H.~Ahn, J.~Alonso-Mora, L.~Carlone, M.~Cap, Y.~F. Chen,
  C.~Choi, J.~Dusek, Y.~Fang \emph{et~al.}, ``Duckietown: an open, inexpensive
  and flexible platform for autonomy education and research,'' in \emph{2017
  IEEE International Conference on Robotics and Automation (ICRA)}.\hskip 1em
  plus 0.5em minus 0.4em\relax IEEE, 2017, pp. 1497--1504.

\bibitem{rubenstein2012kilobot}
M.~Rubenstein, C.~Ahler, and R.~Nagpal, ``Kilobot: A low cost scalable robot
  system for collective behaviors,'' in \emph{2012 IEEE International
  Conference on Robotics and Automation}.\hskip 1em plus 0.5em minus
  0.4em\relax IEEE, 2012, pp. 3293--3298.

\bibitem{rubenstein2015aerobot}
M.~Rubenstein, B.~Cimino, R.~Nagpal, and J.~Werfel, ``Aerobot: An affordable
  one-robot-per-student system for early robotics education,'' in \emph{2015
  IEEE International Conference on Robotics and Automation (ICRA)}.\hskip 1em
  plus 0.5em minus 0.4em\relax IEEE, 2015, pp. 6107--6113.

\bibitem{robinette2009}
P.~Robinette, R.~Meuth, R.~Dolan, , D.~Wunsch, R.~E. Solutions, LLC, Rolla, and
  M.~USA, ``Labrat: Miniature robot for students, researchers, and hobbyists,''
  in \emph{The 2009 IEEE/RSJ International Conference on Intelligent Robots and
  Systems October 11-15, 2009 St. Louis, USA}.\hskip 1em plus 0.5em minus
  0.4em\relax IEEE, 2009.

\bibitem{kernbach2011swarmrobot}
S.~Kernbach, ``Swarmrobot.org -- open-hardware microrobotic project for
  large-scale artificial swarms,'' \emph{arXiv preprint arXiv:1110.5762}, 2011.

\bibitem{cole2012stem}
R.~K. Cole, ``Stem outreach with the boe-bot,'' in \emph{Robots in K-12
  Education: A New Technology for Learning}.\hskip 1em plus 0.5em minus
  0.4em\relax IGI Global, 2012, pp. 245--265.

\bibitem{riedo2013thymio}
F.~Riedo, M.~Chevalier, S.~Magnenat, and F.~Mondada, ``Thymio ii, a robot that
  grows wiser with children,'' in \emph{2013 IEEE Workshop on Advanced Robotics
  and its Social Impacts}.\hskip 1em plus 0.5em minus 0.4em\relax IEEE, 2013,
  pp. 187--193.

\bibitem{mclurkin2014robot}
J.~McLurkin, A.~McMullen, N.~Robbins, G.~Habibi, A.~Becker, A.~Chou, H.~Li,
  M.~John, N.~Okeke, J.~Rykowski \emph{et~al.}, ``A robot system design for
  low-cost multi-robot manipulation,'' in \emph{2014 IEEE/RSJ international
  conference on intelligent robots and systems}.\hskip 1em plus 0.5em minus
  0.4em\relax IEEE, 2014, pp. 912--918.

\bibitem{wilson2016pheeno}
S.~Wilson, R.~Gameros, M.~Sheely, M.~Lin, K.~Dover, R.~Gevorkyan, M.~Haberland,
  A.~Bertozzi, and S.~Berman, ``Pheeno, a versatile swarm robotic research and
  education platform,'' \emph{IEEE Robotics and Automation Letters}, vol.~1,
  no.~2, pp. 884--891, 2016.

\bibitem{stager2017scaled}
A.~Stager, L.~Bhan, A.~Malikopoulos, and L.~Zhao, ``A scaled smart city for
  experimental validation of connected and automated vehicles,'' \emph{arXiv
  preprint arXiv:1710.11408}, 2017.

\bibitem{marquez-sanchez_construction_2019}
C.~Márquez-Sánchez, M.~Antonio-Cruz, J.~Sandoval-Gutiérrez, V.~E.
  Quiroz-Velázquez, and C.~A. Merlo-Zapata, ``Construction of a {Low}-cost
  {Wheeled} {Mobile} {Robot} for {Testing} {Automatic} {Control}
  {Techniques},'' in \emph{2019 {International} {Conference} on {Mechatronics},
  {Electronics} and {Automotive} {Engineering} ({ICMEAE})}.\hskip 1em plus
  0.5em minus 0.4em\relax IEEE, 2019, pp. 102--107.

\bibitem{pickem_robotarium_2017}
D.~Pickem, P.~Glotfelter, L.~Wang, M.~Mote, A.~Ames, E.~Feron, and
  M.~Egerstedt, ``The robotarium: {A} remotely accessible swarm robotics
  research testbed,'' in \emph{2017 {IEEE} {International} {Conference} on
  {Robotics} and {Automation} ({ICRA})}.\hskip 1em plus 0.5em minus 0.4em\relax
  IEEE, 2017, pp. 1699--1706.

\bibitem{liniger2014}
A.~Liniger, A.~Domahidi, and M.~Morari, ``Optimization-based autonomous racing
  of 1:43 scale {RC} cars,'' \emph{Optimal Control Applications and Methods},
  jul 2014.

\bibitem{gonzales2016}
J.~Gonzales, F.~Zhang, K.~Li, and F.~Borrelli, ``Autonomous drifting with
  onboard sensors,'' in \emph{Proceedings of the 13th International Symposium
  on Advanced Vehicle Control (AVEC)}, 2016.

\bibitem{karaman2017}
S.~Karaman, A.~Anders, M.~Boulet, J.~Connor, K.~Gregson, W.~Guerra, O.~Guldner,
  M.~Mohamoud, B.~Plancher, R.~Shin \emph{et~al.}, ``Project-based,
  collaborative, algorithmic robotics for high school students: Programming
  self-driving race cars at mit,'' in \emph{Integrated STEM Education
  Conference (ISEC)}.\hskip 1em plus 0.5em minus 0.4em\relax IEEE, 2017.

\bibitem{okelly2019}
M.~{O'Kelly}, V.~{Sukhil}, H.~{Abbas}, J.~{Harkins}, C.~{Kao}, Y.~{Vardhan
  Pant}, R.~{Mangharam}, D.~{Agarwal}, M.~{Behl}, P.~{Burgio}, and
  M.~{Bertogna}, ``{F1/10: An Open-Source Autonomous Cyber-Physical
  Platform},'' \emph{arXiv e-prints}, p. arXiv:1901.08567, Jan 2019.

\bibitem{hyldmar2019}
N.~{Hyldmar}, Y.~{He}, and A.~{Prorok}, ``{A Fleet of Miniature Cars for
  Experiments in Cooperative Driving},'' \emph{arXiv e-prints}, p.
  arXiv:1902.06133, Feb 2019.

\bibitem{bemani_cyber-physical_2020}
A.~Bemani and N.~Bjorsell, ``Cyber-{Physical} {Control} of {Indoor}
  {Multi}-vehicle {Testbed} for {Cooperative} {Driving},'' \emph{arXiv preprint
  arXiv:2006.04421}, 2020.

\bibitem{reiter2017}
M.~Reiter, M.~Wehr, F.~Sehr, A.~Trzuskowsky, R.~Taborsky, and D.~Abel, ``The
  {IRT}-buggy {\textendash} vehicle platform for research and education,''
  \emph{{IFAC}-{PapersOnLine}}, jul 2017.

\bibitem{vedder_low-cost_2018}
B.~Vedder, J.~Vinter, and M.~Jonsson, ``A low-cost model vehicle testbed with
  accurate positioning for autonomous driving,'' \emph{Journal of Robotics},
  vol. 2018, 2018, publisher: Hindawi.

\bibitem{xray}
\BIBentryALTinterwordspacing
XRAY, ``{M18 PRO LiPo},'' Website, 12 2010, viewed 31 October 2019. [Online].
  Available:
  \url{https://www.teamxray.com/teamxray/products/proddesc.php?prod_id=3622&kategoria=0&catName=XRAY%20M18%20PRO%20LiPo}
\BIBentrySTDinterwordspacing

\bibitem{scheffe2020}
P.~Scheffe, M.~Kloock, A.~Derks, J.~Maczijewski, B.~Alrifaee, and
  S.~Kowalewski, ``Networked and autonomous model-scale vehicles for
  experiments in research and education,'' 2020, {IFAC World Congress},
  accepted.

\bibitem{kloock2020}
M.~Kloock, P.~Scheffe, I.~T{\"u}lleners, J.~Maczijewski, B.~Alrifaee, and
  S.~Kowalewski, ``Vision-based real-time indoor positioning system for
  multiple vehicles,'' \emph{arXiv preprint arXiv:2002.05755}, 2020.

\bibitem{meek1992clothoid}
D.~Meek and D.~Walton, ``Clothoid spline transition spirals,''
  \emph{Mathematics of computation}, vol.~59, no. 199, pp. 117--133, 1992.

\bibitem{althoff2017commonroad}
M.~Althoff, M.~Koschi, and S.~Manzinger, ``{CommonRoad: Composable benchmarks
  for motion planning on roads},'' in \emph{2017 IEEE Intelligent Vehicles
  Symposium (IV)}.\hskip 1em plus 0.5em minus 0.4em\relax IEEE, 2017, pp.
  719--726.

\bibitem{bender2014lanelets}
P.~Bender, J.~Ziegler, and C.~Stiller, ``{Lanelets: Efficient map
  representation for autonomous driving},'' in \emph{2014 IEEE Intelligent
  Vehicles Symposium Proceedings}.\hskip 1em plus 0.5em minus 0.4em\relax IEEE,
  2014, pp. 420--425.

\bibitem{volker2019verification}
M.~V{\"o}lker, M.~Kloock, L.~Rabanus, B.~Alrifaee, and S.~Kowalewski,
  ``Verification of cooperative vehicle behavior using temporal logic,''
  \emph{IFAC-PapersOnLine}, vol.~52, no.~8, pp. 99--104, 2019.

\bibitem{alrifaee2016coordinated}
B.~Alrifaee, F.-J. He{\ss}eler, and D.~Abel, ``Coordinated non-cooperative
  distributed model predictive control for decoupled systems using graphs,''
  \emph{IFAC-PapersOnLine}, vol.~49, no.~22, pp. 216--221, 2016.

\bibitem{kloock2019distributed2}
M.~Kloock, P.~Scheffe, S.~Marquardt, J.~Maczijewski, B.~Alrifaee, and
  S.~Kowalewski, ``Distributed model predictive intersection control of
  multiple vehicles,'' in \emph{2019 IEEE Intelligent Transportation Systems
  Conference (ITSC)}.\hskip 1em plus 0.5em minus 0.4em\relax IEEE, 2019, pp.
  1735--1740.

\bibitem{kloock2019networked}
M.~Kloock, P.~Scheffe, L.~Botz, J.~Maczijewski, B.~Alrifaee, and S.~Kowalewski,
  ``Networked model predictive vehicle race control,'' in \emph{2019 IEEE
  Intelligent Transportation Systems Conference (ITSC)}.\hskip 1em plus 0.5em
  minus 0.4em\relax IEEE, 2019, pp. 1552--1557.

\bibitem{henzinger2001giotto}
T.~A. Henzinger, B.~Horowitz, and C.~M. Kirsch, ``{Giotto: A time-triggered
  language for embedded programming},'' in \emph{International Workshop on
  Embedded Software}.\hskip 1em plus 0.5em minus 0.4em\relax Springer, 2001,
  pp. 166--184.

\bibitem{pardo2003omg}
G.~Pardo-Castellote, ``{OMG data-distribution service: Architectural
  overview},'' in \emph{23rd International Conference on Distributed Computing
  Systems Workshops, 2003. Proceedings.}\hskip 1em plus 0.5em minus 0.4em\relax
  IEEE, 2003, pp. 200--206.

\bibitem{omg2019whois}
{Object Management Group}, ``Who's {U}sing {D}{D}{S}?''
  \url{https://www.omgwiki.org/dds/who-is-using-dds-2/}, 2019, [Online].

\bibitem{furst2016autosar}
S.~F{\"u}rst and M.~Bechter, ``Autosar for connected and autonomous vehicles:
  The autosar adaptive platform,'' in \emph{2016 46th annual IEEE/IFIP
  international conference on Dependable Systems and Networks Workshop
  (DSN-W)}.\hskip 1em plus 0.5em minus 0.4em\relax IEEE, 2016, pp. 215--217.

\bibitem{quigley2009ros}
M.~Quigley, K.~Conley, B.~Gerkey, J.~Faust, T.~Foote, J.~Leibs, R.~Wheeler, and
  A.~Y. Ng, ``{ROS: an open-source Robot Operating System},'' in \emph{ICRA
  workshop on open source software}, vol.~3, no. 3.2.\hskip 1em plus 0.5em
  minus 0.4em\relax Kobe, Japan, 2009, p.~5.

\bibitem{mqtt}
I.~B. M.~C. (IBM) and Eurotech, \emph{MQTT V3.1 Protocol Specification}, 2010.

\bibitem{alrifaee2017networked}
B.~Alrifaee, ``Networked model predictive control for vehicle collision
  avoidance: Vernetzte modellbasierte pr{\"a}diktive regelung zur
  kollisionsvermeidung von fahrzeugen,'' Ph.D. dissertation, RWTH Aachen
  University, 2017.

\bibitem{kloock2019distributed}
M.~Kloock, L.~Kragl, J.~Maczijewski, B.~Alrifaee, and S.~Kowalewski,
  ``Distributed model predictive pose control of multiple nonholonomic
  vehicles,'' in \emph{2019 IEEE Intelligent Vehicles Symposium (IV)}.\hskip
  1em plus 0.5em minus 0.4em\relax IEEE, 2019, pp. 1620--1625.

\end{thebibliography}

\begin{acronym}
    \acro{CPM Lab}{Cyber-Physical Mobility Lab}
    \acro{UDSSC}{University of Delaware Scaled Smart City} 
    \acro{IPS}{Indoor Positioning System}
    \acro{HLC}{High-Level Controller}
    \acro{MLC}{Mid-Level Controller}
    \acro{LLC}{Low-Level Controller}
    \acro{LCC}{Lab Control Center} 
    \acro{IMU}{Inertial Measurment Unit}
    \acro{MPC}{Model Predictive Control}
    \acro{PCB}{Printed Circuit Board} 
    \acro{DDS}{Data Distribution Service}
    \acro{ROS}{Robot Operating System}
    \acro{XiL}{X-in-the-Loop}
    \acro{MIT}{Massachusetts Institute of Technology}
\end{acronym}

\end{document}